\documentstyle[12pt,epsf]{article}

\def\Ar{\rightarrow}

\def\a{\alpha}

\def\n{\nu}

\def\th{\theta}

\def\l{\lambda}

\begin{document}
\vspace{1 cm}
\centerline{\LARGE \bf Deviation of Neutrino Mixing}
\vskip 0.5 cm
\centerline{\LARGE \bf from Bi-maximal}
\vskip 1.5 cm
\centerline{{\large \bf Carlo  Giunti $^1$}
\renewcommand{\thefootnote}{\fnsymbol{footnote}}
\footnote[1]{E-mail address: giunti@to.infn.it, \ 
 Homepage: http://www.to.infn.it/~giunti}
{\large \bf  \quad and \quad Morimitsu Tanimoto  $^2$}
\renewcommand{\thefootnote}{\fnsymbol{footnote}}
\footnote[2]{E-mail address: tanimoto@muse.sc.niigata-u.ac.jp}
 }
\vskip 0.8 cm
\centerline{$^1$ \it{INFN, Sezione di Torino and
Dipartimento di Fisica Teorica, Universit\`a di Torino,}}
\centerline{\it{Via P. Giuria 1, I--10125 Torino, Italy}}
\vskip 1 cm
 \centerline{ $^2$ \it{Department of Physics, Niigata University,
 Ikarashi 2-8050, 950-2181 Niigata, JAPAN}}
\vskip 2.2 cm
\centerline{\bf ABSTRACT}\par
\vskip 1 cm
We have studied  how observables of
the neutrino mixing matrix  can link up with the ones in the quark
sector.
The deviation from the bi-maximal flavor mixing is parameterized
using a $3\times 3$ unitary matrix.
The neutrino mixings are
investigated supposing this unitary matrix
to be hierarchical like the quark mixing matrix.
We obtain the remarkable prediction  $|U_{e3}| \geq 0.03$
from the experimentally allowed range $\tan^2\theta_{\rm sol} = 0.24
\sim 0.89$.
The CP violation in neutrino oscillations is expected to be very small.
\newpage
Recent data from the Super-Kamiokande
\cite{SKam,SKamsolar}
and SNO (Sudbury Neutrino Observatory)
\cite{SNO}
experiments
have provided model independent evidences in favor of oscillations
of atmospheric and solar neutrinos.
The disappearance of atmospheric $\nu_\mu$'s
measured in the Super-Kamiokande experiment
\cite{SKam}
has been confirmed by the
MACRO \cite{MACRO} and Soudan 2 \cite{Soudan} experiments,
and by the K2K long-baseline experiment \cite{K2K}.
The Super-Kamiokande \cite{Fukuda:2000np}
and
MACRO \cite{MACRO} atmospheric neutrino data
favor the $\n_\mu \Ar \nu_\tau$
process.
The solar neutrino data of the
Super-Kamiokande
\cite{SKamsolar}
and SNO
\cite{SNO}
experiments show in a model independent way
that $\n_e \Ar \nu_{\mu,\tau}$
transitions take place.
This evidence agrees with the
comparison of the
Solar Standard Model \cite{SSM}
predictions with the data of the other solar neutrino experiments
(Homestake, GALLEX, SAGE, GNO \cite{{othersolarexp}}).
The global analysis of all solar neutrino data
in terms of
$\n_e \Ar \nu_{\mu,\tau}$
oscillations
\cite{global,Sol}
favor strongly the Large Mixing Angle (LMA) MSW solution
\cite{MSW}.

These experimental results
indicate that neutrinos are massive and mixed particles \cite{MNS,Po}
and the flavor mixing of neutrinos is bilarge,
\textit{i.e.} close to bimaximal \cite{bimax}.
This means that
the neutrino flavor mixing is very different from the one in the quark
sector.
It is therefore important to investigate how the observables of the
neutrino mixing matrix \cite{FT} can link up with those
in the quark sector \cite{CKM}.

In the recent experimental data, the neutrino flavor mixings deviate
from the bimaximal flavor mixing as follows \cite{Sol,Atm}:
\begin{eqnarray}
&&\sin^2 2 \theta_{\rm atm} > 0.83 \quad   ( 99\ \% \  C.L.)\ ,
\nonumber \\
&&\tan^2\theta_{\rm sol} = 0.24 \sim 0.89\quad  ( 99.73\ \% \  C.L.) \ .
\label{data}
\end{eqnarray}
One may consider seriously the deviation  \cite{Expansion}
from the bimaximal flavor mixing  \cite{bimax}.

In this paper, we discuss the deviation from
the bimaximal flavor mixing of neutrinos
by linking it up phenomenologically with the quark flavor mixing.
In our naive understanding it is natural that the charged lepton mass
matrix has a structure similar to the quarks mass matrices.
On the other hand, the neutrino mass matrix has special structure, like the
possibility to be Majorana particles. Therefore the neutrino mixing matrix
is the very different from the CKM matrix.
In the standpoint of this naive understanding, 
the deviation from the bimaximal  links up with the quark mixing.

Let us consider the bimaximal flavor mixing as follows:
\begin{eqnarray}
   \nu_\a  =  U^{(0)}_{\a i} \nu_i \ ,
\end{eqnarray}
\noindent where
\begin{eqnarray}
U^{(0)}= \left ( \matrix{\frac{1}{\sqrt{2}}& \frac{1}{\sqrt{2}} & 0\cr
                -\frac{1}{2} &  \frac{1}{2}& \frac{1}{\sqrt{2}} \cr
 \frac{1}{2} & -\frac{1}{2} &  \frac{1}{\sqrt{2}} \cr } \right ) \ .
\end{eqnarray}
\noindent This bimaximal flavor mixing is supposed to be guaranteed by
a flavor symmetry
\footnote{For example,
  one may  consider $L_e - L_\mu - L_\tau$ symmetry \cite{symmetry}.}
, although  we do not discuss such symmetry in this paper.

One  can parametrize the deviation  $U^{(1)}$ in
$\nu_\a =  [ U^{(1)^\dagger} U^{(0)}]_{\a i} \nu_i$  as
follows
\footnote{We take  $U^{(1)^\dagger}$ in order to compare with quark
mixings.}
:
\begin{equation}
U^{(1)} = \left (\matrix{ c_{13} c_{12} & c_{13} s_{12} &  s_{13} e^{-i
\phi}\cr
  -c_{23}s_{12}-s_{23}s_{13}c_{12}e^{i \phi} &
c_{23}c_{12}-s_{23}s_{13}s_{12}e^{i \phi} &
                       s_{23}c_{13} \cr
  s_{23}s_{12}-c_{23}s_{13}c_{12}e^{i \phi} &
-s_{23}c_{12}-c_{23}s_{13}s_{12}e^{i \phi} &
                       c_{23}c_{13} \cr} \right ) \ ,
\label{Mix}
\end{equation}
\noindent   where  $s_{ij}\equiv \sin{\theta_{ij}}$ and $c_{ij}\equiv
\cos{\theta_{ij}}$ denote the mixing angles  in the bimaximal basis and
$\phi$ is the CP violating Dirac phase.
The mixings $s_{ij}$ are expected to be small since these are deviations
from the bimaximal mixing.
Here, the Majorana phases are absorbed in the neutrino mass eigenvalues.

Let us assume the mixings  $s_{ij}$ to be  hierarchical like the ones
 in the quark sector, $s_{12}\gg s_{23}\gg s_{13}$.
Then, taking the leading contribution due to $s_{12}$, we have
\begin{eqnarray}
|U_{e1}|\simeq \frac{1}{\sqrt{2}} \left (
c_{12}+\frac{1}{\sqrt{2}}s_{12} \right ) \ , \quad
|U_{e2}|\simeq \frac{1}{\sqrt{2}} \left (
c_{12}-\frac{1}{\sqrt{2}}s_{12} \right ) \ , \quad
|U_{e3}|\simeq  \frac{1}{\sqrt{2}} s_{12}  \ ,
\end{eqnarray}
which lead to
\begin{equation}
\tan^2 \theta_{\rm sol}
=
\left( \frac{ c_{12} - \frac{1}{\sqrt{2}}s_{12} }{ c_{12} +
\frac{1}{\sqrt{2}}s_{12} } \right)^2
=
1 - 2 \sqrt{2} s_{12}
+ \mathrm{O}(s_{12}^2)
\,.
\end{equation}
Thus, the solar neutrino mixing is somewhat reduced due to $s_{12}$.
Taking the limits in Eq.(\ref{data}), we get the allowed region
$s_{12}=0.04 \sim 0.43$
\footnote{The  $\mathrm{O}(s_{12}^2)$ terms are taken in order
to estimate the upper bound because those becomes important in the case
of $s_{12}\geq 0.3$.}
.
On the other hand, the experimental upper bound
$|U_{e3}| < 0.2$
obtained from
the results of the CHOOZ experiment \cite{Chooz}
(see Ref.\cite{Bilenky:tw})
gives the limit $s_{12}< 0.28$.
In conclusion,  we get the allowed region
\begin{equation}
  s_{12}=0.04 \sim 0.28 \ ,
\label{s12-bounds}
\end{equation}
which implies
\begin{equation}
|U_{e3}| = 0.03 \sim 0.2
\ .
\label{Ue3}
\end{equation}
Let us emphasize the lower bound for $|U_{e3}|$,
which implies that $|U_{e3}|$
could be measured in the JHF-Kamioka long-baseline neutrino oscillation
experiment
\cite{JHF-Itow,JHF-Aoki},
which has a planned sensitivity of
$|U_{e3}| \simeq 0.04$ at 90\% CL
in the first phase with the Super-Kamiokande detector
and
$|U_{e3}| < 10^{-2}$
in the second phase with the Hyper-Kamiokande detector
\cite{JHF-Itow}.

Next, taking the leading term due to $s_{23}$ and neglecting  $s_{13}$,
we have
\begin{equation}
|U_{\mu 3}|\simeq \frac{1}{\sqrt{2}}c_{12} (c_{23}-s_{23} ) \ , \qquad
|U_{\tau 3}|\simeq \frac{1}{\sqrt{2}} (c_{23}+s_{23} ) \  \ , \
\end{equation}
which give
 $\sin \theta_{\rm atm}$  and $\cos \theta_{\rm atm}$ as follows:
\begin{equation}
\sin^2 \theta_{\rm atm}
=
\frac{|U_{\mu3}|^2}{1-|U_{e3}|^2}
\,,
\qquad
\cos^2 \theta_{\rm atm}
=
\frac{|U_{\tau3}|^2}{1-|U_{e3}|^2} \ .
\label{sc-atm}
\end{equation}
Then,  $\sin^2 2 \theta_{\rm atm}$ is given as
\begin{equation}
\sin^2 2 \theta_{\rm atm}
\simeq
\left[ 1 - s_{12}^2 \left( 1 - (c_{23}-s_{23})^2 \right) \right] \left(
1- 2 s_{23}^2 \right)^2 = 1- \mathrm{O}(s_{12}^4 \sim s_{23}^2) \ .
\label{eqatm}
\end{equation}
Since we have $s_{12}^4 \leq  6 \times 10^{-3}$
from the upper bound in Eq.~(\ref{s12-bounds}),
we predict in practice
\begin{equation}
\sin^2 2 \theta_{\rm atm} = 1 \, .
\label{atm-prediction}
\end{equation}
Thus, the quark-like mixing of $U^{(1)}$ is nicely consistent with
the experimental data.

The CP violation  originates from the phase $\phi$ in $U^{(1)}$.
Keeping $s_{13}$ in the expression of $U_{e3}$, we get
\begin{equation}
\arg\ [U_{e3}]\simeq -\arctan\left [\frac{s_{13}\sin{\phi}}{s_{12}}\right ]\ ,
\label{CP}
\end{equation}
which is the CP violating phase in the standard parametrization 
\footnote{The mixing matrix $U^{(1)^\dagger} U^{(0)}$
can be reduced to the standard form with real
$U_{e1}$, $U_{e2}$, $U_{\mu 3}$, $U_{\tau 3}$
through a rephase of the charged lepton and neutrino fields
that does not change the phase of $U_{e3}$.}
.
 This phase is very small as far as $s_{12}\gg s_{13}$.
Let us estimate the Jarlskog invariant as a measure of CP violation
\cite{Ja}:
\begin{equation}
J = \mathrm{Im}\left[U_{e2}^* U_{e3} U_{\mu2} U_{\mu3}^* \right] \ ,
\label{J}
\end{equation}
which  is written as
\begin{equation}
J
=
\frac{1}{4\sqrt{2}}
\,
c_{13} s_{13}
\left( c_{23}^2 - s_{23}^2 \right)
\left[
\left( c_{12}^2 - s_{12}^2 \right)
\left( c_{23} + s_{23} \right)
\sin\phi
+
c_{12} s_{12} s_{13}
\left( c_{23} - s_{23} \right)
\sin2\phi
\right]
\, .
\label{Jour}
\end{equation}
If we assume the hierarchy of mixing angles
\begin{equation}
s_{13} \ll s_{23} \ll s_{12} \ll 1
\,,
\label{hierarchy}
\end{equation}
as in the quark sector,
the leading contribution to $J$ is given by
\begin{equation}
J
\simeq
\frac{1}{4\sqrt{2}}
\,
s_{13}
\,
\sin\phi
\,.
\label{Japprox}
\end{equation}
\noindent
If $s_{13}$ and $\phi$ are fixed,
one can quantify the smallness of the CP violation
comparing $J$ with its upper limit \cite{Dunietz:1987yt}
\begin{equation}
J \leq \frac{1}{6\sqrt{3}}
\, .
\label{Jmax}
\end{equation}

Let us present numerical predictions of the mixings and 
the CP violating phase.
If the deviation is comparable to the flavor mixing of the quark sector,
the Wolfenstein parametrization is useful  \cite{Wolfen}:
\begin{eqnarray}
U^{(1)} = \left ( \matrix{1- \frac{1}{2}\l^2 & \l & A \l^3 (\rho -
i\eta)\cr
\cr
                -\l  &1- \frac{1}{2}\l^2 & A\l^2 \cr
\cr
  A \l^3 (1-\rho - i\eta) & -  A\l^2 &  1  \cr } \right ) \ ,
\label{para}
\end{eqnarray}
\noindent
where  $\l$, $A$, $\rho$ and $\eta$ are independent of ones
 in the quark sector.

In order to estimate the neutrino mixings, we try to take the same
values of the
quark mixings. Putting typical values of the CKM matrix elements
\cite{CKM},
\begin{equation}
\l=0.22 \ ,  \quad A=0.83 \ , \quad   \rho=0.2 \ , \quad \eta=0.4 \ ,
\label{Wo}
\end{equation}
\noindent we predict  the neutrino mixing matrix
$U  =  U^{(1)\dagger} U^{(0)}$ as
\begin{eqnarray}
| \ U \ | = \left ( \matrix{0.80 & 0.58 & 0.15\cr
                0.35  & 0.66 & 0.66 \cr
                0.48  & 0.48 & 0.74 \cr } \right ) \ ,
\end{eqnarray}
which leads to
\begin{equation}
\sin^2 2 \theta_{\rm atm}= 0.99  \ , \qquad \tan^2 \theta_{\rm sol}=
0.45\ .
\end{equation}
\noindent
These predictions are  nicely consistent with the experimental bounds in
Eq.(\ref{data}).
The solar neutrino mixing is reduced due to $s_{12}$, while
the atmospheric neutrino mixing is not reduced as seen in
Eq.(\ref{eqatm}).
The prediction $|U_{e3}|=0.15$ is not much below the experimental
upper bound $\simeq 0.2$.
Therefore will not be difficult to test this prediction
in the near future,
for example in the JHF-Kamioka neutrino experiment
\cite{JHF-Itow,JHF-Aoki}.

Let us  present  the $\lambda$ dependence of our results.
We  show in Fig.1 the predictions for
$\tan^2 \th_{\rm sol}$ and
$\sin^2 2\th_{\rm atm}$
as functions of $|U_{e3}|$,
obtained varying $\lambda$ from $0$ to $0.28$
and
keeping the values of the other parameters
given in Eq.(\ref{Wo}).
One can see that $|U_{e3}|$ is predicted to be larger than about $0.03$
(see Eq.(\ref{Ue3})),
under the condition that $\tan^2\theta_{\rm sol}\leq 0.89$.

We can predict  the amount of CP violation in neutrino oscillations.
If we assume that
\begin{equation}
s_{13} \sim s_{12}^3
\,,
\label{s13}
\end{equation}
as in the quark sector,
from the upper bound for
$s_{12}$ obtained from solar neutrino data
($s_{12} < 0.28$),
we obtain
\begin{equation}
J_{\mathrm{max}}
\sim
4 \times 10^{-3}
\,,
\label{Jmax-est}
\end{equation}
\noindent
which is about $4\times10^{-2}$ times smaller than
the maximum possible value of $J$
in Eq.(\ref{Jmax}).
It is interesting to compare the value of $J_{\mathrm{max}}$
that we have obtained in Eq.~(\ref{Jmax-est}) with
the maximum value of $J$
that is possible in a general
quasi-bimaximal mixing scheme,
\textit{i.e.}
a scheme with
\begin{equation}
|U_{e1}|
\simeq
|U_{e2}|
\simeq
|U_{\mu3}|
\simeq
|U_{\tau3}|
\simeq
\frac{1}{\sqrt{2}}
\,,
\qquad
|U_{e3}|
\ll
1
\,.
\label{quasi-bimaximal}
\end{equation}
In such scheme $J$ is approximately given by
\begin{equation}
J^{\mathrm{(bimax)}}
\simeq
\frac{1}{4}
\,
\mathrm{Im}[U_{e3}^*]
\,,
\label{J-bim}
\end{equation}
and its maximum possible value is
\begin{equation}
J^{\mathrm{(bimax)}}_{\mathrm{max}}
\simeq
\frac{1}{4}
\,
|U_{e3}|
\,.
\label{Jmax-bim}
\end{equation}
Taking  the bound
\begin{equation}
|U_{e3}|
< 0.2
\,,
\label{Ue3max}
\end{equation}
obtained from CHOOZ data,
we have
\begin{equation}
J^{\mathrm{(bimax)}}_{\mathrm{max}}
\simeq
5 \times 10^{-2}
\,.
\label{Jmax-bim-est}
\end{equation}
Comparing Eqs.~(\ref{Jmax-est}) and (\ref{Jmax-bim-est})
one can see that the maximum value of the Jarlskog invariant $J$
in our scheme
is about an order of magnitude smaller than
the maximum value of $J$
in a generic quasi-bimaximal scheme.
Therefore, the  CP violation seems too small to be measured in JHF
\cite{JHF-Itow,JHF-Aoki},
but maybe
it can be measured in a neutrino factory \cite{Albright:2000xi}.
On the other hand, the Majorana phases are not constrained,
but unfortunately they are not measurable in neutrino oscillation
experiments.

We summarize as follows.
 We have studied  how observables of the
 neutrino mixing matrix  can link up with the ones in the quark sector.
The deviation from the bimaximal flavor mixing  is parametrized
 by  a $3\times 3$ unitary matrix.  Supposing that this unitary matrix
is similar to the quark mixing matrix, we predict the neutrino mixings,
 which are consistent with the experimental data.
The element $U_{e3}$ of the neutrino mixing matrix
is predicted to be larger than  $0.03$
by using the experimental bound on the solar neutrino mixing.
When more solar neutrino data will be available in the near future,
a more precise prediction will be given for $U_{e3}$.
For instance, if we use  $\tan^2\theta_{\rm sol}\leq  0.58$
(90\% C.L. at present), we predict  $|U_{e3}|\geq 0.11$.
The violation of CP is predicted to be very small.
Thus, the measurements of  the solar neutrino mixing and $U_{e3}$
\cite{JHF-Itow,JHF-Aoki}
will present a crucial test for our scheme.

\vskip 1  cm

 This research is supported by the Grant-in-Aid for Science Research,
 Ministry of Education, Science and Culture, Japan(No.12047220).

\newpage

  \newpage
\begin{figure}
\epsfxsize=14 cm
\centerline{\epsfbox{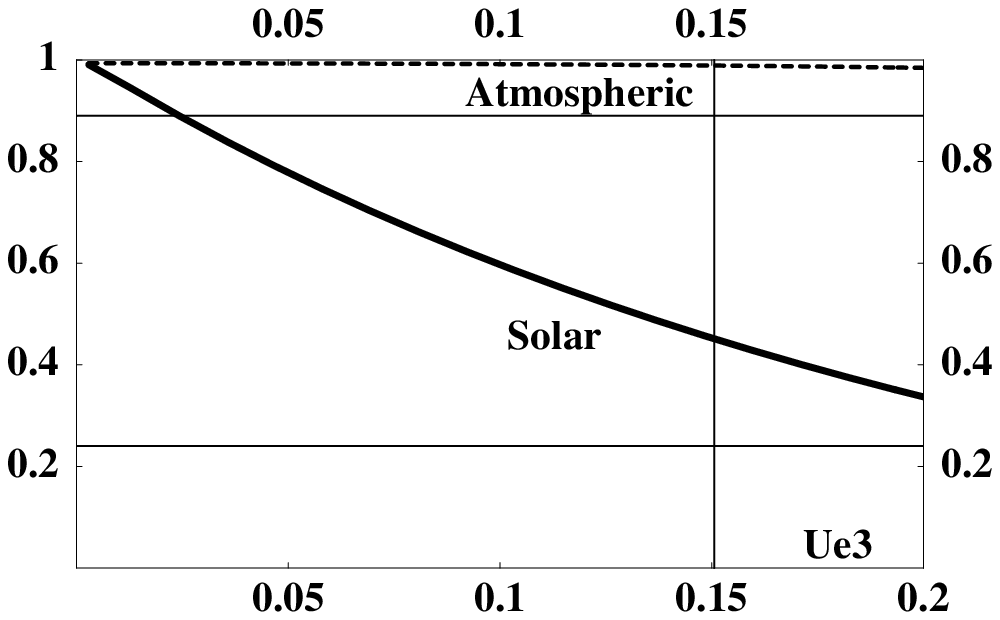}}
\caption{Predictions in the $|U_{e3}|-\tan^2 \th_{\rm sol}$ plane and
$|U_{e3}|-\sin^2 2\th_{\rm atm}$ plane.
The thick solid curve  corresponds to $\tan^2 \th_{\rm sol}$,
while the dashed one to  $\sin^2 2\th_{\rm atm}$.
Horizontal lines delimit the experimental allowed regions for solar
neutrinos.
The parameter $\lambda$ is varied from  0 to  0.28.
The vertical line around $|U_{e3}|=0.15$
corresponds to the result in the case of $\lambda=0.22$.}
\end{figure}

\end{document}